# Modeling Curved Carbon Fiber Composite (CFC) Structures in the Transmission-Line Modeling (TLM) Method

Xuesong Meng, Phillip Sewell, *Senior Member, IEEE*, Sendy Phang, Ana Vukovic, *Member, IEEE*, Trevor M. Benson, *Senior Member, IEEE*

*Abstract*—A new embedded model for curved thin panels is developed in the Transmission Line Modeling (TLM) method. In this model, curved panels are first linearized and then embedded between adjacent 2D TLM nodes allowing for arbitrary positioning between adjacent node centers. The embedded model eliminates the necessity for fine discretization thus reducing the run time and memory requirements for the calculation. The accuracy and convergence of the model are verified by comparing the resonant frequencies of an elliptical cylinder formed using carbon fiber composite (CFC) materials with those of the equivalent metal cylinder. Furthermore, the model is used to analyze the shielding performance of CFC airfoil NACA2415.

## I. INTRODUCTION

IN recent years, carbon fiber composite (CFC) materials have received considerable attention in the aerospace and aircraft industries, due to their high strength-to-weight ratio and ease of fabrication. However, their lower conductivity results in lower shielding effectiveness compared to that of metals. A considerable amount of work [1]-[3] has been done on investigating the shielding effectiveness of CFC materials [4], but most of this has been focused on planar CFC panels. However, there are a number of instances where the CFC panels are curved, such as cylinder-shaped CFC structures used in aircraft fuselages [5]-[6].

As the CFC structures are usually very thin, a smaller mesh size is needed to discretize them within a computational electromagnetic simulation, resulting in longer run time and more computational overheads. One possible way for modelling fine features is to use a non-uniform mesh, which allows the small mesh only apply to the areas where fine features are present. For example, a multi-grid mesh has been reported in the numerical Finite Difference Time Domain (FDTD) method in [7]-[10] and the Transmission Line Modeling Method (TLM) in [11]-[14], and a multi-level Octree mesh has been used in the CST microwave studio software [15]. However, the number of time steps is determined by the smallest mesh size, so the computational burden is still heavy.

An alternative method is to embed them in the mesh without actually discretizing them. The mesh size, and consequently the time step, is chosen according to the frequency of interest and not the smallest feature in the model. This is done in the numerical FDTD in [2], where the thin layer is first modeled using frequency domain surface impedance formulations, which are then transformed to the time domain by applying vector fitting (VF) procedures.

Similarly, the embedding planar panels in the numerical Transmission Line Modeling (TLM) method has been reported in [16] where an internal boundary condition that allows for arbitrary positioning of embedded models in the TLM mesh has been presented. The approach uses the Prony method [17] to extract frequency domain transfer functions from known frequency domain data samples, and then uses a bi-linear transform to obtain the time domain response. The disadvantage of this approach is that the Prony method requires a correct choice of the number of poles for good approximation and needs to deal with inverse matrix problems. To avoid using the Prony method, an embedded model for a thin planar layer in the TLM method has recently been reported [18], where the analytical expansions of cotangent and cosecant functions in the admittance matrix of the thin layer are used to efficiently obtain its frequency response.

Modeling of curved boundaries using a structured mesh is a more demanding problem that requires fine mesh sizes to resolve the curvature of objects and in order to reduce the numerical noise that arises from piece-wise discretization. Additional constraints on the mesh size are placed when the thickness is much smaller than other physical dimensions of the curved object. In this paper the approach presented in [18] is extended to modeling thin curved panels, which can be composed of ideal or lossy dielectric materials. Embedding of the curved thin panels in the TLM mesh is done firstly, by approximating the panels using a piece-wise linearization and secondly, embedding the linearized segments between adjacent

Manuscript received xx, 2014. This work was supported by China Scholarship Council.
The authors are with the George Green Institute for Electromagnetics Research, University of Nottingham, Nottingham, NG7 2RD, UK. (email:Xuesong.meng@nottingham.ac.uk)

nodes allowing for arbitrary placement between the nodes. A three-layer planar stack is introduced to allow for arbitrary model placement between TLM nodes. The frequency response of the three-layer stack is transformed to the time domain by using an inverse Z transform and general digital filter technique as in [18], which results in a slight modification of the TLM algorithm.

The structure of the paper is as follows: In section II a short overview of the two-dimensional (2D) TLM method is given together with the description of the embedded model for curved CFC structures. The validation of the model is presented in Section III and the application of the model in modeling the shielding effectiveness of a CFC airfoil with and without gaps is presented in section IV. Section V outlines the main conclusions of the paper.

## II. Curved CFC Model in the TLM

In this section, the basis of the 2D TLM method is outlined, together with the approach for embedding thin curved panels within the square TLM mesh.

The TLM method [19] is a time-domain electromagnetic simulation technique, which is based on the analogy between the electromagnetic fields and electric circuits. It can be used to model a range of materials, including frequency-dependent [20], anisotropic [21] and nonlinear [22] materials. TLM methods based on both structured and unstructured meshes, as [23]-[24], have been reported.

In the 2D TLM model, free space is modeled by using two different nodes: the series node shown in Fig.1 (a) and the shunt node shown in Fig.1 (b). As shown in Fig.1, the series and shunt TLM nodes are represented by four ports, 1, 2, 3 and 4. At each port the total voltage is represented as a sum of incident voltages, e.g. $V_1^i, V_2^i, V_3^i$ and $V_4^i$, and reflected voltages, e.g. $V_1^r, V_2^r, V_3^r$ and $V_4^r$. $Z_{TL}$ is the characteristic impedance of each transmission line. In the series node, $Z_{TL} = Z_0/\sqrt{2}$, where $Z_0$ is the characteristic impedance of free space, while in the shunt node, $Z_{TL} = \sqrt{2}Z_0$.

The TLM algorithm consists of three distinct steps namely, initialization, scattering and connection. The total voltage on link lines is a summation of incident and reflected voltages. In the initiation phase the incident voltages on TLM nodes are obtained. In the scattering phase the voltages reflected from the node are calculated and in the connection phase the reflected voltages on the link lines become incident on the link lines of neighboring nodes. This process is repeated as required. For the full details of the TLM algorithm the reader is referred to [19].

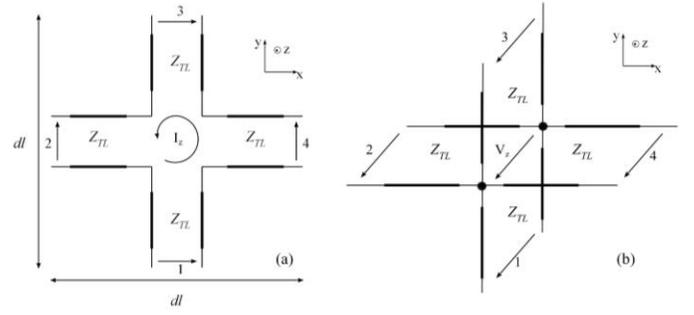

Fig.1 (a) the series TLM node (b) the shunt TLM node in the 2D TLM model.

Fig.2 shows the schematic of a curved thin film, represented by the solid curve (green), positioned within a coarse mesh of size *dl*, represented by dashed lines (black). The solid cross lines (red) represent the transmission link lines along which the voltages travel, and the point of their intersection is defined as the TLM node center. The curved thin film is firstly approximated by linear piece-wise segments, represented by dash-dot lines (blue), each of which can be viewed as a planar thin film. The linearization is done by connecting the crossing points of the arc and the link lines of the nodes. The crossing point can be either exactly between two nodes (point A on Fig.2 (b)) or can split the transmission line of a node at an arbitrary position (point B in Fig.2 (b)). The curved panel thus needs to be embedded at each crossing point. If the arc is defined by a function $y = f(x)$, then the position of points A and B in an arbitrary node $(n_x, n_y)$ can be expressed as $(n_x \cdot dl, f(n_x \cdot dl))$ and $(f^{-1}(n_y \cdot dl), n_y \cdot dl)$, respectively.

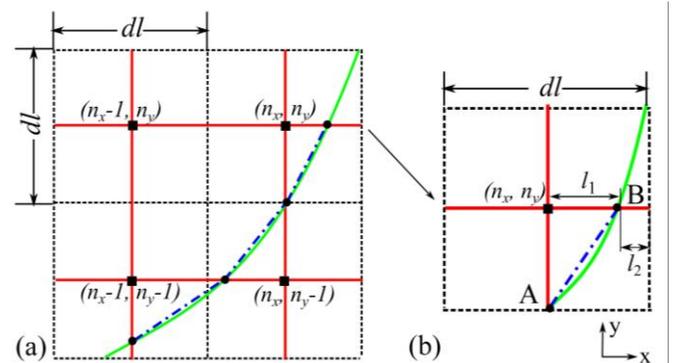

Fig.2 (a) A curved CFC structure embedded between 2D TLM nodes (b) The enlarged node $(n_x, n_y)$.

At the crossing point A, the curved panel is modeled and embedded as a transmission line positioned centrally between the two adjacent nodes with coordinates $(n_x, n_y)$ and $(n_x, n_y - 1)$, which is done by modifying the TLM's connection process as in [18]. However, at the crossing point B, the curved panel splits the transmission link line at the right side of the node $(n_x, n_y)$ into two segments of lengths $l_1$ and $l_2$. In this case, the whole section of transmission line together with the section of the curved panel is modeled and embedded as a three-layer stack whereby the curved panel is sandwiched between two sections of transmission lines of lengths $l_1$ and $l_2$ as shown in Fig.3.

Fig.3 shows the three-layer stack embedded between two 2D series nodes $(n_x, n_y)$ and $(n_x+1, n_y)$, where the two shaded layers represent the transmission lines of the node and the middle layer (blue) represents the curved panel. In the figure, $_xV_4^i$ and $_xV_4^r$ are the incident and reflected voltages at port 4 of the node $(n_x, n_y)$, while $_{x+1}V_2^i$ and $_{x+1}V_2^r$ are the incident and reflected voltages at port 2 of the node $(n_x+1, n_y)$.

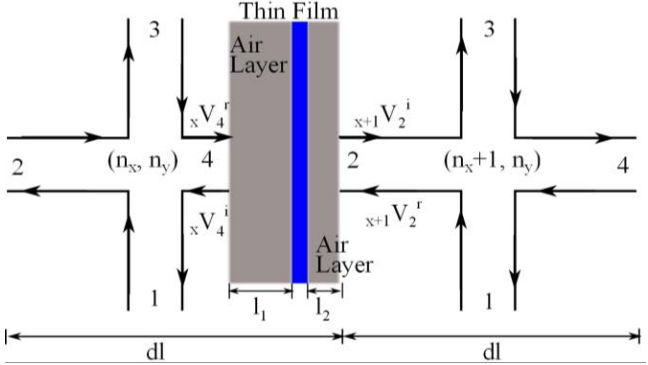

Fig. 3. Two layers of air together with the planar thin film are composed of a three-layer stack embedded between two adjacent 2D nodes.

The transmission line model of this three-layer stack embedded between two TLM nodes is shown in Fig.4. Transmission lines of lengths $l_1$ and $l_2$ are, in this case, made of air but in general can represent any material parameters.

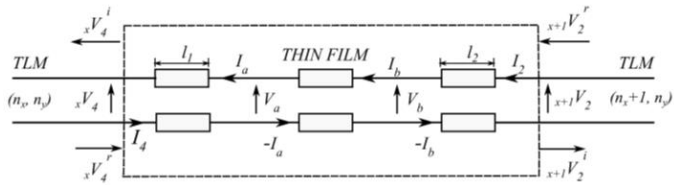

Fig. 4. Transmission line model of the three-layer stack embedded between two TLM nodes.

The admittance matrix of each layer in the three-layer stack can be expressed as [18], [25]

$$\begin{pmatrix} I_4 \\ I_a \end{pmatrix} = \begin{pmatrix} y_{TL} - jY_0 \cot\theta_1 & jY_0 \csc\theta_1 \\ jY_0 \csc\theta_1 & -jY_0 \cot\theta_1 \end{pmatrix} \begin{pmatrix} _xV_4 \\ V_a \end{pmatrix}, \quad (1)$$

$$\begin{pmatrix} -I_a \\ I_b \end{pmatrix} = \begin{pmatrix} -jY_t \cot\theta_t & jY_t \csc\theta_t \\ jY_t \csc\theta_t & -jY_t \cot\theta_t \end{pmatrix} \begin{pmatrix} V_a \\ V_b \end{pmatrix}, \quad (2)$$

$$\begin{pmatrix} -I_b \\ I_2 \end{pmatrix} = \begin{pmatrix} -jY_0 \cot\theta_2 & jY_0 \csc\theta_2 \\ jY_0 \csc\theta_2 & y_{TL} - jY_0 \cot\theta_2 \end{pmatrix} \begin{pmatrix} V_b \\ _{(x+1)}V_2 \end{pmatrix}, \quad (3)$$

where $Y_0 = \sqrt{\varepsilon_0/\mu_0}$ is the admittance of the air layer, and $\theta_1$ and $\theta_2$ are the electrical lengths of the two layers of air, described as

$$\theta_i = \omega l_i \sqrt{\mu_0 \varepsilon_0}, \quad i=1,2 \quad (4)$$

where $\omega$ is the angular frequency.

The electrical length of the thin film that represents the curved panel is $\theta_t = \omega d \sqrt{LC}$ where $d$ is the thickness of the film and $Y_t = \sqrt{C/L}$ is the admittance of the thin film where L and C can be expressed as in [19]

$$L = \mu + \sigma_m/j\omega, \quad C = \varepsilon + \sigma_e/j\omega. \quad (5)$$

Combining equations (1), (2) and (3), the linear matrix equations (6) can be obtained.

The terms of the left hand side of equation (6) are known and are

$$I_4 = 2y_{TL} \cdot _xV_4^r, \quad I_2 = 2y_{TL} \cdot _{(x+1)}V_2^r.$$

The unknown voltages on its right side, $_xV_4, V_a, V_b$ and $_{(x+1)}V_2$, can be solved using an iterative matrix solver based on the Gauss-Seidel method [26]. These solutions are given in the frequency domain and need to be transferred into the time domain to enable time-stepping of the TLM code. This is done by using an inverse Z-transform and digital filter theory.

In order to do that, the cotangent and cosecant functions in equations (1)-(3) are expanded as an infinite summation in the form [27]

$$\cot\theta = \frac{1}{\theta} + 2\theta \sum_{k=1}^{N=\infty} \frac{1}{\theta^2 - k^2\pi^2},$$
$$\csc\theta = \frac{1}{\theta} + 2\theta \sum_{k=1}^{N=\infty} \frac{(-1)^k}{\theta^2 - k^2\pi^2}. \quad (7)$$

where $N$ denotes the number of terms in the expansion and determines the accuracy of the expansion.

With this expansion in place, the solutions of equation (6) i.e. $_xV_4, V_a, V_b$ and $_{(x+1)}V_2$ are first transferred to the s-domain using the transformation $s=j\omega$ and then to the Z-domain using the transformation

$$s = \frac{2}{\Delta t} \frac{1-z^{-1}}{1+z^{-1}}.$$

The final solutions are expressed in the time domain using the inverse Z transform and general digital filter theory as in [18].

$$\begin{pmatrix} I_4 \\ 0 \\ 0 \\ I_2 \end{pmatrix} = \begin{pmatrix} y_{TL} - jY_0 \cot\theta_1 & jY_0 \csc\theta_1 & 0 & 0 \\ jY_0 \csc\theta_1 & -jY_0 \cot\theta_1 - jY_t \cot\theta_t & jY_t \csc\theta_t & 0 \\ 0 & jY_t \csc\theta_t & -jY_0 \cot\theta_2 - jY_t \cot\theta_t & jY_0 \csc\theta_2 \\ 0 & 0 & -jY_0 \csc\theta_2 & y_{TL} - jY_0 \cot\theta_1 \end{pmatrix} \begin{pmatrix} {}_pV_4 \\ V_a \\ V_b \\ {}_{(x+1)}V_2 \end{pmatrix} \quad (6)$$

### III. VALIDATIONS

In this section the proposed model for embedding thin curved panels is used to extract the resonant frequencies of metal and CFC elliptical cylinders. Results obtained are compared against the known analytical values for the metal elliptical cylinder.

In order to investigate the accuracy of the linearization of the curved panel, the resonant frequencies of a metal elliptical cylinder with a major axis $a = 10$ cm and minor axis $b = 6.614$ cm [28] are considered. As the cylinder is metallic the curved panel model shown in Fig. 4 reduces to only one transmission line which is terminated with a metal boundary. An input signal in the form of a delta pulse was launched from a point source located at the point (0.1 m, 0.06 m). The TLM simulation was run for $2 \cdot 10^5$ time steps. Both TE and TM polarizations were considered.

Fig.5 shows the percentage errors between TLM and analytical resonant frequencies for the first six TE and TM modes for different discretizations, represented as the ratio $b/dl$. The analytical values of the metal elliptical cylinder are taken from [28]. Fig.5 indicates that TLM results show good agreement with the analytical ones.

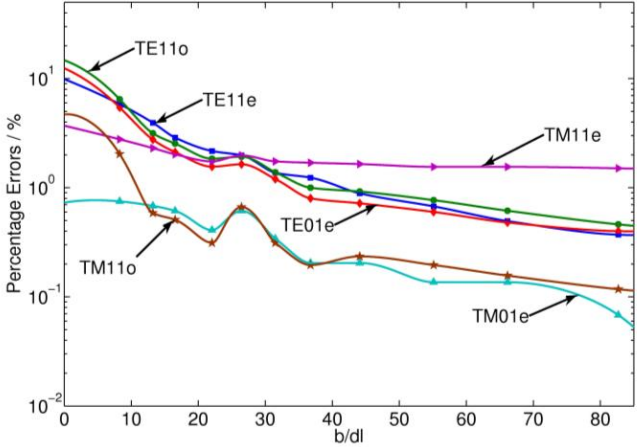

Fig. 5. The percentage errors in the resonant frequencies for TE and TM modes of metal elliptical cylinder.

As said in the reference [29], if the conductivity of the elliptical cylinder wall is finite but very large, the fields inside such structures have the same forms as those in the metal structure. Therefore, the resonant frequencies of the CFC cylinder are compared with those of metal cylinder for verification. Fig.6 compares the relative differences in resonant frequencies of the CFC cylinder and metal cylindrical resonator. The CFC material was modeled as a linear and isotropic medium, characterized by constant effective conductivity and permittivity [1]. The CFC parameters were chosen as in [2] with effective permittivity $\varepsilon_r = 2$, conductivity $\sigma_e = 10000 S/m$ and thickness $d = 1mm$. The CFC structure was embedded in the 2D TLM mesh using the model described in section II. The 2D computation window was set to $40cm \times 26cm$, and terminated with matched boundaries [19]. The number of time steps used in the calculation was $2 \cdot 10^5$.

Fig.6 shows the relative differences in the resonant frequencies of the first six TE and TM modes of the CFC and metal elliptical cylinder for different mesh sizes represented by the $b/dl$ parameter.

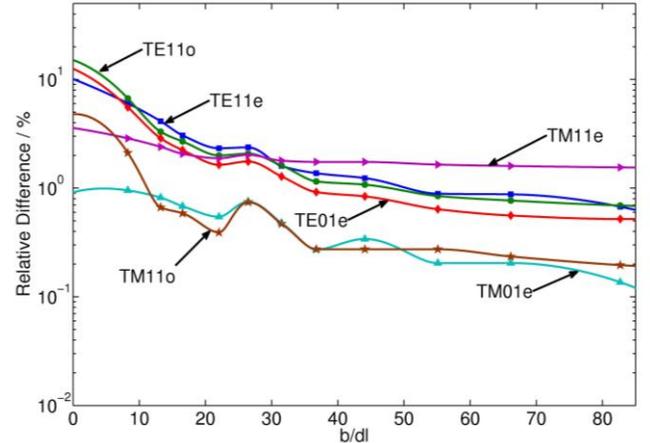

Fig. 6. The relative differences of the resonant frequencies for TE and TM modes in the CFC elliptical cylinder and metal elliptical cylinder.

It can be seen that the relative differences in the even (e) and odd (o) $TE_{11}$ mode resonant frequencies converge to around 0.7% and the relative differences in the even $TE_{01}$ mode resonant frequencies converge to around 0.5% as the mesh size decreases. Fig.6 also indicates that the relative differences in the even $TM_{11}$ mode resonant frequencies converge to around 1.56% and the relative differences in the odd $TM_{11}$ and even TM01 mode resonant frequencies converge to around 0.2% as the mesh size decreases. Fig.6 confirms that the resonant frequencies of the CFC resonator are close to those of the metal resonator, confirming the metal-like properties of the CFC material.

Table I further compares the resonant frequencies of the metallic and CFC elliptical cylinder for the first six modes when the mesh size $dl = 0.8$ mm ($b/dl = 82.7$). It can be seen that the relative differences in the resonant frequencies of the CFC cylinder and metal cylinder are very small and within 1.56%.

TABLE I
The resonant frequencies for the six lowest modes of an elliptical CFC cylinder compared to those of an elliptical metal cylinder

| Modes | Resonant Frequency (GHz) | | Relative Difference (%) |
|---|---|---|---|
| | Metal elliptical cylinder | CFC elliptical cylinder | |
| Even $TE_{11}$ | 0.889 | 0.883 | 0.67 |
| Odd $TE_{11}$ | 1.30 | 1.291 | 0.69 |
| Even $TM_{01}$ | 1.467 | 1.465 | 0.14 |
| Even $TM_{11}$ | 2.124 | 2.091 | 1.56 |
| Even $TE_{01}$ | 2.50 | 2.487 | 0.52 |
| Odd $TM_{11}$ | 2.554 | 2.549 | 0.20 |

## IV. APPLICATIONS

In this section the embedded model is applied to analyze the shielding performance of a CFC airfoil structure. Furthermore, the impact of small gaps in the airfoil structure on the shielding effectiveness is also investigated.

### A. Shielding performance of a CFC airfoil structure

The profile of an airfoil structure is taken from the National Advisory Committee for Aeronautics (NACA) report [30]. An airfoil with the profile NACA2415 from the NACA four-digit series is taken as an example.

In the NACA four-digit series, the first digit specifies the maximum camber ($m$) in percentage of the chord (airfoil length $c$); the second digit indicates the position of the maximum camber along the chord ($p$) in tenths of chord; the last two digits provide the maximum airfoil thickness ($t$) in percentage of chord.

In the example of the airfoil NACA2415, the airfoil length $c$ was chosen as 1m. It has a maximum thickness $t = 0.15$ m with a camber m = 0.02 m located 0.4 m back from the airfoil leading edge. Based on these values, the coordinates for the entire airfoil can be computed using the analytical equations reported in [28]. Fig.7 shows the profile of the airfoil NACA2415 modeled using the TLM method, with a computational window size of $1.2\text{m} \times 0.3\text{m}$ terminated with matched boundary conditions. The parameters of the CFC materials were chosen as before: effective permittivity $\varepsilon_r = 2$, conductivity $\sigma_e = 10000 S/m$, and thickness $d = 1mm$.

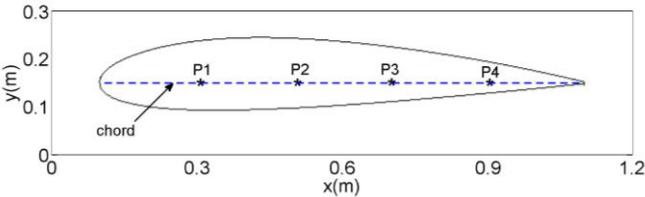

Fig. 7. The CFC airfoil NACA2415 modeled by TLM

In the model, 2D shunt nodes were used to model free space. A TE polarized plane wave was used as an excitation, propagating from $y = 0$ to $y = 0.3$ m. The electric field shielding effectiveness $SE_E$ is defined as in [31]:

$$SE_E = 20 \cdot \log_{10}(|E^{without}/E^{with}|), \quad (8)$$

in which $E^{without}, E^{with}$ are the magnitudes of the electric field component in the same point without and with the shield, respectively.

The magnitude of the electric $E_z$ field is observed at four points along the chord, i.e. P1 (0.3 m, 0.15 m), P2 (0.5 m, 0.15 m), P3 (0.7 m, 0.15 m), and P4 (0.9 m, 0.15 m) as shown in Fig.7, with and without the CFC airfoil. The shielding effectiveness (SE) is computed at these four specific points along the chord. Fig. 8 shows the electric field shielding effectiveness of the CFC airfoil NACA2415 at the points P1, P2, P3 and P4 in the frequency range from 1 GHz to 2 GHz. The TLM mesh size used is $dl = 2mm$. It can be seen that the SE of the airfoil becomes much smaller at certain frequencies due to resonance effects [32]. Fig. 8 also indicates that the SE at the points P1 and P2 are similar in the frequency range from 1 GHz to 2 GHz, while the SE at the point P3 is higher than that at the points P1 and P2 at the frequencies below 1.2 GHz and the SE at the point P4 is much higher than that at the points P1 and P2 at the frequencies below 1.7 GHz. At higher frequencies, the SE is very similar at the four points because the existence of the higher modes contributes to an even distribution of the electric field in the structure.

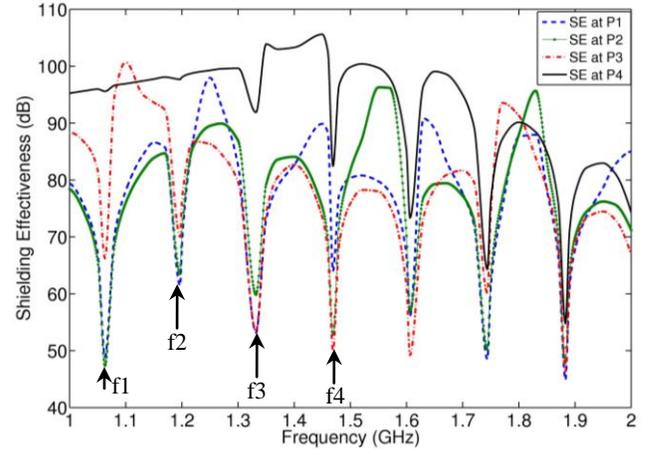

Fig. 8. The electric field shielding effectiveness of the CFC airfoil NACA2415.

To explain the lower SE at the points P1 and P2 at lower frequencies, compared to that at the points P3 and P4, the scattering of the CFC airfoil NACA2415 when illuminated by the TE wave at $f_1 = 1.063 GHz$ is shown in Fig.9. The electric field intensity in the 2D space at the time step of $10^4$ is plotted in dB. As shown in the figure, due to the non-metallic properties of the CFC panel, the electric field penetrates the airfoil and excites the first resonant mode. The center of the resonant mode is near the points P1 and P2, leading to the lower shielding effectiveness at these points at 1.063 GHz. The resonance has little effect on the electric field at the point P4 so the SE at the point P4 is high.

It is emphasized that the field is shown on a logarithmic scale and that the intensity of the field inside the CFC airfoil is small compared to that of the excitation wave, as shown in Fig.9.

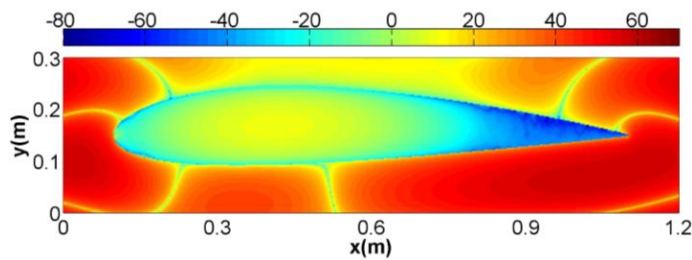

Fig. 9. The scattering of the CFC airfoil NACA2415 upon the 1.063 GHz TE wave illumination. The plot shows electric field intensity on a dB scale.

In order to show the convergence of the proposed model, the first four resonant frequencies of the CFC airfoil (labeled as f1, f2, f3 and f4 in Fig.8) were calculated using different mesh sizes and are shown in Fig.10 as a function of $t/dl$. The figure indicates that as the mesh size decreases, all four resonant frequencies converge.

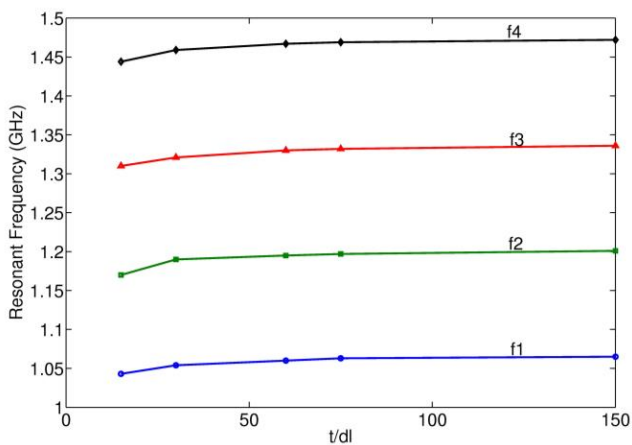

Fig. 10. The first four resonant frequencies of the CFC airfoil NACA2415 against t/dl.

### B. Shielding performance of CFC airfoil structure with gaps

Imperfections on the airfoil, such as gaps, even if very small, can affect the SE of the structure [33]. This section investigates the impact of the gaps on the SE of the CFC airfoil.

The same CFC airfoil NACA2415 is used and the gap is positioned in the downside of the structure at $x=0.9m$. Two examples were chosen, one with a gap of 2 mm and the other with a gap of 6 mm. The excitation of the problem is the same as that for Fig.8. The shielding effectiveness at the points P1 and P4 along the chord of the CFC airfoil is computed and shown in Fig. 11 (a) and (b), respectively, for two different size gaps in the airfoils and compared to the case of no gaps in the range from 1 GHz to 2 GHz.

Fig.11 (a) shows when the frequency is below 1.1 GHz, the gaps in the airfoil do not affect the SE at the point P1. However, as the frequency increases, the SE at the point P1 decreases rapidly for both airfoils with 2 mm and 6 mm gaps compared to the case of no gaps in the airfoil. For example, at the seventh resonant frequency (1.883 GHz), the shielding effectiveness of the airfoils with 2 mm and 6 mm gaps is around 25 dB and 14 dB, respectively, compared to 45 dB with the case of no gaps. This can be explained by the fact that at higher frequencies, the shorter wavelength of the signal results in an increased penetration of fields through the gap, which change the field distribution in the airfoil.

Different from the SE at the point P1, the SE at the point P4 is greatly influenced by the gaps in the airfoil as shown in Fig.11 (b). The SE at the point P4 is drastically reduced in the frequency range from 1 GHz to 2 GHz, for both airfoils with 2 mm and 6 mm gaps compared to the case of no gaps in the airfoil. The reduction of the SE at the point P4 is more prominent at the lower frequency. For example, at 1.063 GHz, the SE at the point P4 of the airfoil with 2 mm and 6 mm gaps is reduced by around 40 dB and 50 dB, respectively, compared to that of the airfoil without gaps.

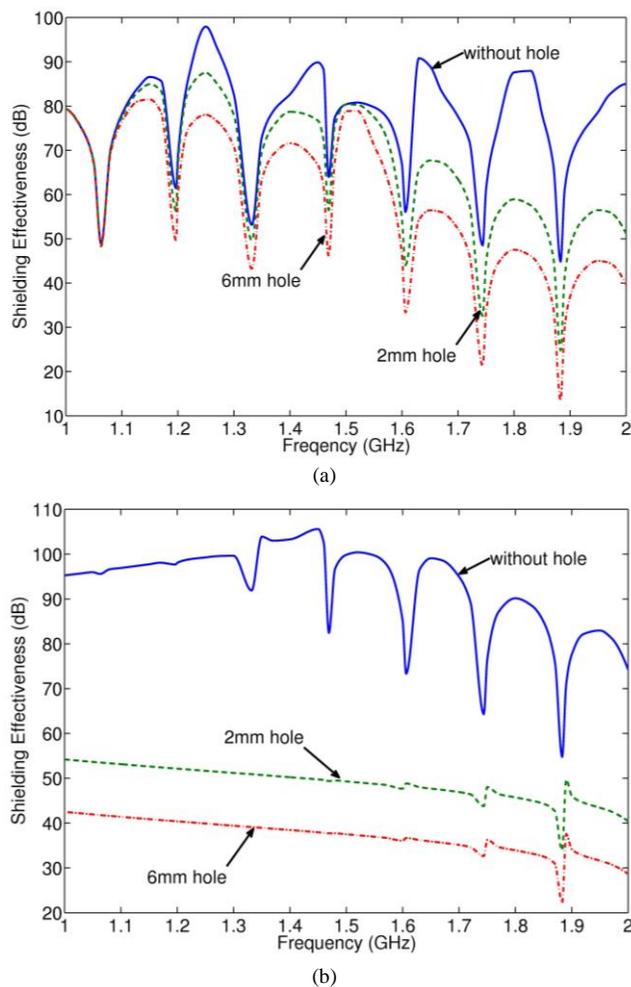

Fig. 11. The shielding effectiveness of the CFC airfoil NACA2415 with no gap, a 2 mm gap and 6 mm gap (a) at the point P1 and (b) at the point P4.

To explain the prominent reduction of the SE at the point P4 at lower frequencies, the scattering of the CFC airfoil NACA2415 with a 2 mm hole under illumination from a TE wave at $f_1 = 1.063 GHz$ is shown in Fig.12. The electric field intensity in the 2D space at the time step of $10^4$ is plotted in dB. It can be seen that the small gap allows the incident field to more readily couple with the inside of the airfoil. Compared to Fig. 9, the field intensity is increased at the tail of the airfoil, resulting in reduced shielding performance of the airfoil at the point P4.

Comparison of Fig.12 and Fig.9 indicates that the field penetrating through the gap does not greatly perturb the resonant field at point P1 at 1.063 GHz, so the SE at that point is not significantly affected by the gap, as seen in Fig.11 (a).

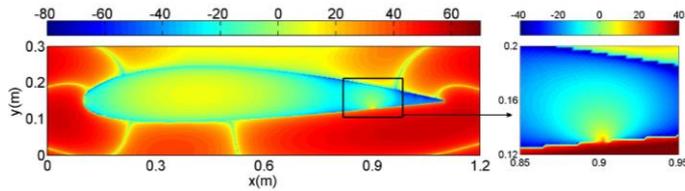

Fig. 12.  The scattering of the CFC airfoil NACA2415 with 2 mm gap in the 1.063 GHz TE wave illumination. The plot shows electric field intensity on a dB scale.

## V. Conclusion

An approach for embedding curved thin panels within a structured TLM mesh is presented in this paper. The method is applied to the case of lossy CFC panels, although any type of dielectric can be included. The implementation is done by firstly, linearizing the curvature of the panel and secondly, representing the panels as three-layer transmission lines to allow for arbitrary positioning of the panel within the mesh.

The model is validated by firstly comparing the resonant frequencies of a metallic 2D elliptical resonator with analytical values to verify the accuracy of the linearization. The difference between the CFC and metallic elliptical resonator has also been obtained confirming good metallic properties of the CFC material. Finally the model is applied to analyze the shielding performance of a CFC airfoil NACA2415 structure. The impact of the small gap in the CFC airfoil on the SE is also reported, showing considerable worsening in the SE performance.

The generality of the approach suggests that extension to 3D problems is possible in future work.

## Acknowledgment

X. Meng would like to thank the China Scholarship Council for their financial support.

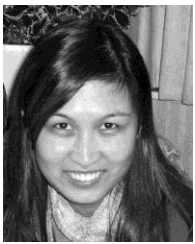

**Xuesong Meng** was born in Hebei, China, in 1986. She received the Master degree in electronic engineering from Beihang University, Beijing, China, in 2010. She is currently working towards the Ph.D. degree in electrical and electronic engineering at the University of Nottingham, Nottingham, U.K.

Her current research interests include numerical modeling of complex materials, optoelectronics and EMC.

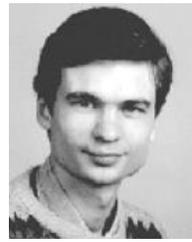

**Phillip Sewell** (M'89-SM'04) was born in London, U.K., in 1965. He received the B.Sc. degree in electrical and electronic engineering (with first-class honors) and Ph.D. degree from the University of Bath, Bath, U.K., in 1988 and 1991, respectively.

From 1991 to 1993, he was a Post-Doctoral Fellow with the University of Ancona, Ancona, Italy. In1993, he became a Lecturer with the School of Electrical and Electronic Engineering, University of Nottingham, Nottingham, U.K. In 2001 and 2005, he became a Reader and Professor of electromagnetics at the University of Nottingham. His research interests involve analytical and numerical modeling of electromagnetic problems with application to opto-electronics, microwaves, and electrical machines.

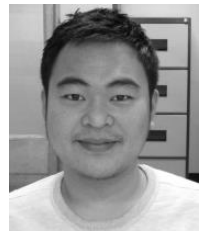

**Sendy Phang** was born in Jakarta, Indonesia in 1988. He received the B.Eng. degree in engineering physics with first class honors, from the Bandung Institute of Technology, Bandung, Indonesia in 2010 and the M.Sc. degree in electromagnetic design from the University of Nottingham in 2011. Since 2012, he has been working at the GGIEMR (George Green Institute for Electromagnetics Research) group at the University of Nottingham as a Ph.D. His research interests include the modeling of dispersive and non-linear optical devices, periodic structures, parity-time symmetric structure in optics and their applications.

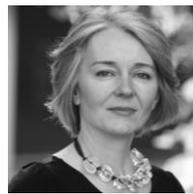

**Ana Vukovic** (M'97) was born in Nis, Yugoslavia, in 1968. She received the Diploma of Engineering degree in electronics and telecommunications from the University of Nis, Nis, Yugoslavia, in 1992, and the Ph.D. degree from the University of Nottingham, Nottingham, U.K., in 2000.

From 1992 to 2001, she was a Research Associate with the University of Nottingham. In 2001, she joined the School of Electrical and Electronic Engineering, University of Nottingham, as a Lecturer. Her research interests are electromagnetics with a particular emphasis on applications in optoelectronics, microwaves, and EMC.

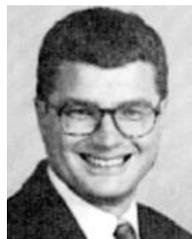

**Trevor M. Benson** received a First Class honors degree in Physics and the Clark Prize in Experimental Physics from the University of Sheffield in 1979, a PhD in Electronic and Electrical Engineering from the same University in 1982 and the DSc degree from the University of Nottingham in 2005.

After spending over six years as a Lecturer at University College Cardiff, Professor Benson moved to The University of Nottingham in 1989. He was promoted to a Chair in Optoelectronics in 1996, having previously been Senior Lecturer (1989) and Reader (1994). Since October 2011 he has been Director of the George Green Institute for Electromagnetics Research at The University of Nottingham. Professor Benson's research interests include experimental and numerical studies of electromagnetic fields and waves with particular emphasis on the theory, modeling and simulation of optical waveguides, lasers and amplifiers, nano-scale photonic circuits and electromagnetic compatibility.

He is a Fellow of the Institute of Engineering Technology (FIET) and the Institute of Physics (FInst.P). He was elected a Fellow of the Royal Academy of Engineering in 2005 for his achievements in the development of versatile design software used to analyze propagation in optoelectronic waveguides and photonic integrated circuits.